\documentclass[aps,journal=jcp,twocolumn,noeprint,superscriptaddress]{revtex4-1}
\usepackage{graphicx}
\usepackage{epstopdf}
\usepackage{bmpsize}
\usepackage{amsmath,amsfonts,amsthm,bm}
\usepackage{bigints}
\usepackage{tikz}
\usetikzlibrary{decorations.pathreplacing}
\usepackage{subcaption}
\usepackage[bookmarks=false]{hyperref}
\hypersetup{colorlinks=true, citecolor=blue, urlcolor=blue, linkcolor=blue}

\makeatletter
\renewcommand{\maketag@@@}[1]{\hbox{\m@th\normalsize\normalfont#1}}%
\makeatother

\newcommand{\ph}{{\phantom a}}
\def\!{\mskip-\thinmuskip}

\graphicspath{{figs/}}

\begin{document}

\title{Nonadiabatic Dynamics of Molecules Interacting with Metal Surfaces: A Quantum-Classical Approach Based on Langevin Dynamics and the Hierarchical Equations of Motion}
\author{Samuel L. Rudge}
\affiliation{Institute of Physics, University of Freiburg, Hermann-Herder-Strasse 3, 79104 Freiburg, Germany}
\email[Corresponding author: Samuel Rudge \\ Email: ]{samuel.rudge@physik.uni-freiburg.de}
\author{Christoph Kaspar}
\affiliation{Institute of Physics, University of Freiburg, Hermann-Herder-Strasse 3, 79104 Freiburg, Germany}
\author{Robin L. Grether}
\affiliation{Institute of Physics, University of Freiburg, Hermann-Herder-Strasse 3, 79104 Freiburg, Germany}
\author{Steffen Wolf}
\affiliation{Institute of Physics, University of Freiburg, Hermann-Herder-Strasse 3, 79104 Freiburg, Germany}
\author{Gerhard Stock}
\affiliation{Institute of Physics, University of Freiburg, Hermann-Herder-Strasse 3, 79104 Freiburg, Germany}
\author{Michael Thoss}
\affiliation{Institute of Physics, University of Freiburg, Hermann-Herder-Strasse 3, 79104 Freiburg, Germany}

\begin{abstract}
\noindent {A novel mixed quantum-classical approach to simulating nonadiabatic dynamics of molecules at metal surfaces is presented. The method combines the numerically exact hierarchical equations of motion approach for the quantum electronic degrees of freedom with Langevin dynamics for the classical degrees of freedom, namely low-frequency vibrational modes within the molecule. The approach extends previous mixed quantum-classical methods based on Langevin equations to models containing strong electron-electron or quantum electronic-vibrational interactions, while maintaining a nonperturbative and non-Markovian treatment of the molecule-metal coupling. To demonstrate the approach, nonequilibrium transport observables are calculated for a molecular nanojunction containing strong interactions.}
\end{abstract}

\maketitle

\section{Introduction}

The dynamics of molecules interacting with metal surfaces is a highly relevant topic in both physics and chemistry, with many technological applications. It covers a broad range of physical scenarios, including the scattering of molecules off surfaces \cite{Tully1990,HeadGordon1995}, reactive and catalytic processes \cite{Saalfrank2006,Brandbyge1995,Kim2002}, and charge transport through STM setups \cite{Persson1997,Saalfrank2003} or molecular nanojunctions \cite{Huang2006,Koch2006b,Schulze2008,Gelbwaser-Klimovsky2018}. Particularly interesting in these systems is understanding how energy is transferred between the continuum of electronic states in the metal and the molecule, and then dispersed among the molecule's vibrational degrees of freedom. Such electronic and vibrational relaxation processes are often nonadiabatic and require theoretical treatments that go beyond the Born-Oppenheimer approximation.

%In quantum transport through molecular nanojunctions, electronic current often couples to the vibrational motion of atomic nuclei. These electronic-vibrational interactions can significantly affect not only the the electronic transport properties, but also the molecule's nuclear dynamics and thus the junction stability. As a result, understanding how energy is transferred between the metallic electrodes and the molecule, and then dispersed among its nuclear degrees of freedom, is crucial for describing many phenomena, such as quantum shuttling \cite{Gorelik1998,Lai2015}, current-induced bond rupture \cite{Erpenbeck2020,Erpenbeck2019,Erpenbeck2018,Sabater2015,Li2015,Li2016}, and vibrational instabilities \cite{Schinabeck2018,Haertle2008,Haertle2011b,Kast2011,Lue2010,Lue2011,Lue2012,Foti2018,Simine2012}. 

The most accurate approaches, such as the multilayer multiconfigurational time-dependent Hartree \cite{Wang2003,Wang2009,Manthe2008,Vendrell2011,Wang2011,Wang2016} or the hierarchical equations of motion (HEOM) \cite{Ke2021,Schinabeck2016,Erpenbeck2019,Tanimura2020,Jin2007,Jin2008} methods, treat all degrees of freedom quantum mechanically. However, they can be numerically challenging to implement, which motivates mixed quantum-classical approaches, in which vibrational degrees of freedom within the molecule move classically on potential energy surfaces but are also coupled to quantum electronic degrees of freedom. There is a diverse range of such approaches, like Ehrenfest dynamics \cite{Bellonzi2016,Subotnik2010,Verdozzi2006,Kartsev2014,Cunningham2015,Stock2005,Dundas2009} or surface hopping \cite{Tully1971,Tully1990,Dou2015a,Dou2015b,Dou2015c,Shenvi2008,Shenvi2009a,Shenvi2009b}. In previous work, for example, we combined the Ehrenfest approach with the HEOM method to simulate reaction dynamics at metal surfaces \cite{Erpenbeck2018}. In this work, we focus exclusively on Langevin dynamics (LD), which incorporates weak nonadiabatic effects on nuclear motion via coupling to electron-hole pairs \cite{HeadGordon1995,Maurer2016}.

In Langevin dynamics, one must first calculate the electronic forces, such as the friction, from a suitable quantum method. Although this can be done from first-principles \cite{Maurer2016}, in recent years, a number of new approaches based on quantum transport methods have arisen \cite{Kershaw2018,Kershaw2020,Preston2020,Preston2023,Dou2015d,Dou2016a}. In Ref.\ \cite{Rudge2023}, for example, two of the authors demonstrated how to calculate the electronic friction within the HEOM formalism, which generalizes methods such as nonequilibrium Green's functions \cite{Mozyrsky2006,Dou2017a,Chen2019a,Chen2019b}, scattering theory \cite{Bode2011,Bode2012,Todorov2011,Todorov2014}, and path integrals \cite{Lue2012,Brandbyge1995,Daligault2007,Hussein2010,Metelmann2011} by including strong interactions within the molecule, all while treating the molecule-metal coupling in a nonperturbative and non-Markovian manner. Consequently, the HEOM approach to electronic forces represents a significant step forward in the range of molecular models available for Langevin dynamics. However, in Ref.\ \cite{Rudge2023}, the authors restricted their investigation exclusively to the forces themselves. 

In this work, to our knowledge for the first time, the electronic forces calculated from HEOM are used in the corresponding Langevin equation, introducing the novel HEOM-LD approach to nonadiabatic dynamics of molecules interacting with metal surfaces. First, a brief introduction to HEOM is presented, alongside a summary of how one obtains electronic forces from it. Next, it is shown how to incorporate these forces into the corresponding Langevin equation and how to then solve for the dynamics. Finally, the new approach is demonstrated via several illustrative examples, including molecular models with strong electron-electron and quantum electronic-vibrational interactions. While the HEOM-LD method is completely applicable to transient molecular dynamics, in this work, all results will focus on one of the most challenging problems in context of open quantum systems, namely calculating steady-state nonequilibrium transport observables in molecular nanojunctions. Alongside these mixed quantum-classical results, numerically exact transport calculations from the fully quantum HEOM are also provided, verifying the accuracy of the HEOM-LD method in appropriate parameter regimes. These quantum calculations are performed via a novel implementation of the HEOM designed to treat low-frequency vibrational modes, which is an extension of the method presented in Ref.\ \cite{Schinabeck2018}.
%
%The first system, a Holstein model, is used to explore the parameter regimes in which it is appropriate to use HEOM-LD. The family of LD approaches are built upon an assumption of weak nonadiabacity, which manifests via a weak electronic-vibrational interaction or a timescale separation between fast electronic degrees of freedom and slow vibrational degrees of freedom. To enforce the latter condition, which is explicitly required for HEOM-LD, contemporary studies usually apply the limit of a large molecule-lead coupling, $\Gamma$, in relation to the effective frequency of the vibrational mode, $\omega$ \cite{Eidelstein2013,Dou2018b,Dou2015c,Dou2015d,Dou2015b,Wilner2014}. In this analysis, the exact quantum calculations allow for a systematic exploration of this assumption, with the surprising conclusion that HEOM-LD is robust at room temperature even when $\Gamma \sim \omega$. This is attributed to the noninteracting nature of the quantum part of the nanosystem once the vibrational mode is treated classically. The next two models introduce interactions to the Holstein model, first an electron-electron interaction via a Coulomb repulsion and then an electronic-vibrational interaction via the addition of a high-frequency quantum vibrational mode. For these models, it is shown that HEOM-LD is actually restricted to the $\Gamma > \omega$ regime. The presence of these quantum interactions also stabilizes the nanojunction, which manifests through periodic negative differential vibrational excitation.

Overall, the example applications of HEOM-LD presented in this work opens up calculations of molecules interacting with metal surfaces to models that were previously inaccessible. For example, many molecules have multiple vibrational modes participating in the transport with very different frequencies, for which fully quantum treatments can be prohibitively expensive and mixed quantum-classical treatments are inaccurate. An approach such as HEOM-LD, in contrast, allows one to systematically separate those parts of the nanosystem requiring a quantum treatment and those that will submit to a classical one. 

%Additionally, the fully quantum HEOM used to benchmark the calculations is restricted to harmonic vibrational modes linearly coupled to the electronic degrees of freedom. For realistic molecular models, in which there are nonlinear electronic-vibrational couplings, low-frequency anharmonic modes, and strong interactions, HEOM-LD is a much more efficient option.

The paper is structured as follows. The model describing a molecular nanojunction is introduced in Section \ref{sec: Model}. Section \ref{sec: Quantum transport theory} contains the quantum transport theory used for the numerically exact comparisons, while in Section  \ref{sec: Electronic friction and Langevin dynamics} an overview of HEOM-LD is presented. This includes a summary of how one calculates the electronic forces and a description of the numerical algorithm used to propagate the Langevin equation. The results of the direct comparison between quantum HEOM and HEOM-LD are contained in Section \ref{sec: Results} and the conclusions are presented in Section \ref{sec: Conclusion}.
% Although most theory is delegated to other papers, any relevant derivations can be found in Appendices \ref{app: Limiting behavior of high-frequency vibrational excitation in 1L2M model}.

Throughout the paper, units are used such that $\hbar = e = m_{e} = 1$.

\section{Model}\label{sec: Model}

In this section, the general model describing nonadiabatic dynamics of molecules interacting with metal surfaces is introduced. While the theory is broadly applicable to any of the scenarios mentioned in the introduction, such as scattering or desorption problems, the specific focus in Sec.\ \ref{sec: Results} will be on nonequilibrium transport in molecular nanojunctions. With this in mind, the electronic degrees of freedom within the metal will be referred to as the leads, with Hamiltonian  $H_{\text{leads}}$. These are coupled to the molecule, $H_{\text{mol}}$, via the molecule-lead interaction, $H_{\text{mol-leads}}$. The total Hamiltonian is 
\begin{align}
H & = H_{\text{mol}} + H_{\text{leads}} + H_{\text{mol-leads}}. \label{eq: total Hamiltonian}
\end{align}

The molecular model contains a series of $N_{\text{el}}$ spin-independent electronic states linearly coupled to $N_{\text{vib}}$ harmonic vibrational modes, with Hamiltonian 
\begin{align}
H_{\text{mol}} = & \sum_{m=1}^{N_{\text{el}}} \varepsilon_{m} d^{\dag}_{m}d^{\ph}_{m} + \sum_{m < m'} U_{mm'}d^{\dag}_{m}d^{\ph}_{m}d^{\dag}_{m'}d^{\ph}_{m'} \nonumber \\
& +  \sum_{m=1}^{N_{\text{el}}}\sum_{i=1}^{N_{\text{vib}}}\lambda_{i}\sqrt{2m_{i}\omega_{i}}\:\hat{x}_{i}  d^{\dag}_{m}d^{\ph}_{m} \nonumber \\
& + \sum_{i = 1}^{N_{\text{vib}}}\left(\frac{\hat{p}^{2}_{i}}{2m_{i}} + \frac{1}{2} m_{i}\omega_{i}^{2}\hat{x}^{2}_{i}\right). \label{eq: general system Hamiltonian}
\end{align}
The $m$th electronic level is described by its annihilation and creation operators, $d^{\ph}_{m}$ and $d^{\dag}_{m}$, with some Coulomb repulsion, $U_{mm'}$, between electrons. The vibrational degrees of freedom, meanwhile, are described by coordinates $\hat{\boldsymbol{x}} = \{\hat{x}_{1},\dots,\hat{x}_{N_{\text{vib}}}\}$ and momenta $\hat{\boldsymbol{p}} = \{\hat{p}_{1},\dots,\hat{p}_{N_{\text{vib}}}\}$. These degrees of freedom can represent center-of-mass motion of atomic nuclei or bond vibrations. Vibrational operators have been written with explicit operator notation to distinguish between the quantum and classical treatments. Under a transformation to dimensionless coordinates, $\hat{x}_{i} \rightarrow \hat{x}_{i}\sqrt{m_{i}\omega_{i}}$ and $\hat{p}_{i} \rightarrow \hat{p}_{i}/\sqrt{m_{i}\omega_{i}}$, the molecular Hamiltonian simplifies to
\begin{align}
H_{\text{mol}} = & \sum_{m=1}^{N_{\text{el}}} \varepsilon^{\ph}_{m} d^{\dag}_{m}d^{\ph}_{m} + \sum_{m < m'} U_{mm'}d^{\dag}_{m}d^{\ph}_{m}d^{\dag}_{m'}d^{\ph}_{m'} \nonumber \\
& + \sum_{m=1}^{N_{\text{el}}}\sum_{i=1}^{N_{\text{vib}}}\lambda_{i}\sqrt{2}\hat{x}_{i}  d^{\dag}_{m}d^{\ph}_{m} + \sum_{i = 1}^{N_{\text{vib}}}\frac{\omega_{i}}{2}\left(\hat{p}^{2}_{i} + \hat{x}^{2}_{i}\right). \label{eq: general system Hamiltonian transformed}
\end{align}
In transport approaches that treat the vibrational degrees of freedom quantum mechanically, it will prove useful to introduce bosonic annihilation and creation operators, $b^{\ph}_{i}$ and $b^{\dag}_{i}$, which are related to the $i$th vibrational mode via 
\begin{align}
\hat{x}_{i} & =  \frac{1}{\sqrt{2}}\left(b^{\ph}_{i} + b^{\dag}_{i}\right) \hspace{0.5cm} \text{ and } \hspace{0.5cm} \hat{p}_{i} = \frac{i}{\sqrt{2}}\left(b^{\ph}_{i} - b^{\dag}_{i}\right).
\end{align}

Although the molecular Hamiltonian in Eq.\eqref{eq: general system Hamiltonian} assumes harmonic oscillators and linear electronic-vibrational couplings, the HEOM-LD approach is not restricted in complexity. The theory presented in Sec.\ \ref{sec: Electronic friction and Langevin dynamics} is completely applicable to molecules containing multiple spin-dependent electronic states, anharmonic vibrational modes with nonlinear couplings, and even nonadiabatic electronic-vibrational interactions. 

The left, $\alpha = L$, and right, $\alpha = R$, metallic leads are modeled with noninteracting electrons and are assumed to be initially in local equilibrium, defined by temperatures $T_{\alpha}$ and chemical potentials $\mu_{\alpha}$. In the junction-type setups considered within Sec.\ \ref{sec: Results}, a voltage bias is applied symmetrically between the left and right leads: $e\Phi = \mu_{L} - \mu_{R}$ with $\mu_{L} = -\mu_{R} = e\Phi /2$. The leads' Hamiltonian is 
\begin{align}
H_{\text{leads}} & = \sum_{\alpha \in \{L,R\}}\sum_{k} \varepsilon^{\ph}_{k \alpha} c^{\dag}_{k \alpha}c^{}_{k \alpha}, \label{eq: general bath Hamiltonian}
\end{align}
where the operators $c^{\ph}_{k\alpha}$ and $c^{\dag}_{k\alpha}$ annihilate and create an electron in lead $\alpha$ with energy $\varepsilon_{k\alpha}$, respectively. These states are linearly coupled to the $m$th electronic state in the molecule via the molecule-lead interaction Hamiltonian, 
\begin{align}
H_{\text{mol-lead}} & = \sum_{m}\sum_{\alpha,k} V_{m,k\alpha}\left(c^{\dag}_{k\alpha}d^{\ph}_{m} + d^{\dag}_{m}c^{}_{k\alpha}\right), \label{eq: general system-bath Hamiltonian}
\end{align}
which introduces $V_{m,k\alpha}$ as the coupling strength between electronic level $m$ in the molecule and electronic state $k$ in lead $\alpha$. The molecule-lead coupling is also characterized by the level-width function of lead $\alpha$, 
\begin{align}
\Gamma_{m,\alpha}(\omega) & = 2\pi  \sum_{k\alpha} \: V^{2}_{m,k\alpha} \delta(\omega - \varepsilon_{k\alpha}). 
\end{align}
In this work, all results are calculated for a Lorentzian density of states, such that
\begin{align}
\Gamma_{m,\alpha}(\omega) & = \Gamma_{m,\alpha}\frac{W^{2}}{(\omega - \mu_{\alpha})^{2} + W^{2}},
\end{align} 
with a bandwidth of $W = 10\text{ eV}$. Throughout the paper, the parameter $\Gamma_{m,\alpha} = 2\pi |V_{m,\alpha}|^{2}$ will be referred to as the molecule-lead coupling. Here, the $V_{m,\alpha}$ now represent a constant coupling strength between state $m$ in the molecule and lead $\alpha$, with the dependence on states in the leads, $k$, being described solely by the Lorentzian density of states.

In the HEOM approach, it is necessary to describe the influence of the leads on the time-evolution of the molecular degrees of freedom. For the noninteracting leads and linear molecule-lead coupling introduced in Eqs.\eqref{eq: general bath Hamiltonian} and \eqref{eq: general system-bath Hamiltonian}, this influence is entirely described by two-time lead-correlation functions \cite{Tanimura2006,Jin2007},
\begin{align}
C^{\sigma}_{m,\alpha}(t - \tau) & = V^{2}_{m,\alpha}\sum_{k\alpha} \text{Tr}_{\text{leads}}\left\{ c^{\sigma}_{k\alpha}(t)c^{\bar{\sigma}}_{k\alpha}(\tau)\rho_{\text{leads}}(0)\right\}. \label{eq: two-time bath-correlation function 1}
\end{align}
In Eq.\eqref{eq: two-time bath-correlation function 1}, the compact notation $\sigma = \pm$ and $\bar{\sigma} = \mp$ with $c^{-}_{k\alpha} = c^{\ph}_{k\alpha}$ and $c^{+}_{k\alpha}  = c^{\dag}_{k\alpha}$ has been introduced, alongside the interaction picture of the leads,  $c^{\sigma}_{k\alpha}(t) = e^{i H_{\text{leads}}t}c^{\sigma}_{k\alpha}e^{-i H_{\text{leads}}t}$. Additionally, it has been implicitly assumed that the initial state of the junction factorizes, 
\begin{align}
\rho_{\text{total}}(0) & = \rho(0)\rho_{\text{leads}}(0), \label{eq: factorized initial state}
\end{align}
where $\rho(0)$ is the initial state of the molecule and $\rho_{\text{leads}}(0)$ is a Gibbs state:
\begin{align}
\rho_{\text{leads}}(0) & = \frac{e^{-\left(H_{\text{L}} - \mu_{\text{L}}\right)/k_{B}T_{\text{L}}}}{\text{Tr}_{\text{L}}\left[e^{-\left(H_{\text{L}} - \mu_{\text{L}}\right)/k_{B}T_{\text{L}}}\right]}\frac{e^{-\left(H_{\text{R}} - \mu_{\text{R}}\right)/k_{B}T_{\text{R}}}}{\text{Tr}_{\text{R}}\left[e^{-\left(H_{\text{R}} - \mu_{\text{R}}\right)/k_{B}T_{\text{R}}}\right]}.
\end{align}
Computationally, the lead-correlation functions can be difficult objects to include. In order to obtain a closed set of equations in the HEOM approach, for example, one expands the $C^{\sigma}_{m,\alpha}(t - \tau)$ as a series of exponential functions,
\begin{align}
C^{\sigma}_{m,\alpha}(t) & \approx |V_{m,\alpha}|^{2} \sum_{\ell = 0}^{\ell_{\max}} \eta_{\alpha,\sigma,\ell,m} e^{-\kappa_{\alpha,\sigma,\ell,m}t}, \label{eq: exponential decomposition}
\end{align}
which is generated via a pole decomposition of the Fermi-Dirac function using the AAA algorithm \cite{Dan2023,Xu2022,Nakatsukasa2018}. All results in this work were calculated at $T = 300$K, such that the decomposition of the bath-correlation functions converged at $\ell_{\max} = 8$.

\section{Quantum Transport Theory}\label{sec: Quantum transport theory}

This section introduces the numerically exact HEOM approach as the quantum transport method underpinning all results in the paper. This includes both the standard formulation, in which molecular vibrations are treated within the molecular density matrix, and the reservoir formulation, in which some vibrational degrees of freedom in the molecule are treated as an extra bath. To finish the section, a discussion on how to calculate relevant transport observables, such as the electric current and vibrational excitation, is presented. 

\subsection{Hierarchical Equations of Motion}\label{subsec: HEOM}

The HEOM transport method, which derives from the Feynman-Vernon influence functional, incorporates the effect of the leads on the time-evolution of the molecular density matrix by coupling the molecular density matrix, $\rho(t)$, to a series of auxiliary density operators (ADOs), $\rho^{(n)}_{\bm{j}}(t)$, in a hierarchical fashion. Since the complete derivation is lengthy, and the approach used in this work closely follows previous formulations, this section will just present a brief overview and leave the details to Refs. \cite{Tanimura1989,Tanimura2006,Haertle2015,Jin2007,Jin2008,Xiao2009,Yan2014,Wenderoth2016,Tanimura2020,Ye2016,Schinabeck2018}. 

Considering the setup and underlying assumptions introduced in Sec.\ \ref{sec: Model}, the equation of motion for an $n$th tier ADO, $\rho^{(n)}_{\bm{j}}(t)$, is 
\begin{align}
\frac{\partial}{\partial t}\rho^{(n)}_{\bm{j}}(t) & = -i\left[H_{\text{S}},\rho^{(n)}_{\bm{j}}\right]  - \left(\sum_{r = 1}^{N} \kappa^{}_{j_{r}}\right)\rho^{(n)}_{\bm{j}} \nonumber \\
& \:\:\:\:\:  - i\sum_{r=1}^{n} (-1)^{n-r}\mathcal{C}_{j_{r}}\rho^{(n-1)}_{\bm{j}^{-}} - i\sum_{j} \mathcal{A}^{\bar{\sigma}}_{\alpha,m} \rho^{(n+1)}_{\bm{j}^{+}}. \label{eq: HEOM Final}
\end{align}
Each $n$th tier ADO is described by an $n$-dimensional vector of super-indices, $\bm{j} = \left(j_{n}, \dots , j_{1}\right)$, where each index $j_{r} = \{\alpha_{j_{r}},\sigma_{j_{r}},\ell_{j_{r}},m_{j_{r}}\}$ can be thought of as a mapping of the essentially continuous interactions between lead and molecule to a series of finite ones. The reduced density matrix of the molecule is simply the $0$th-tier ADO, $\rho^{(0)}(t) = \rho(t)$. Within the HEOM, there are also two more relevant super-indices, $\bm{j}^{-}  = \left(j_{n},\dots , j_{r+1},j_{r-1},\dots, j_{1}\right)$ and $\bm{j}^{+} = \left(j,j_{n},\dots, j_{1}\right)$, which are formed by removing and adding a molecule-lead interaction from and to an ADO, respectively. They relate directly to the superoperators $\mathcal{C}_{j_{r}}$ and $\mathcal{A}^{\bar{\sigma}}_{\alpha,m}$, which couple different tiers of the hierarchy,
\begin{align}
\mathcal{C}_{j_{r}}\rho^{(n)}_{\bm{j}}(t) & = V^{}_{m,\alpha}\left(\eta^{}_{j_{r}}d^{\sigma}_{m}\rho^{(n)}_{\bm{j}}(t) - (-1)^{n}\eta^{*}_{\bar{j}_{r}}\rho^{(n)}_{\bm{j}}(t)d^{\sigma}_{m}\right), \label{eq: coupling down superoperator} \\
\mathcal{A}^{\bar{\sigma}}_{\alpha,m}\rho^{(n)}_{\bm{j}}(t) & = V^{}_{m,\alpha} \left(d^{\bar{\sigma}}_{m}\rho^{(n)}_{\bm{j}}(t) + (-1)^{n}\rho^{(n)}_{\bm{j}}(t)d^{\bar{\sigma}}_{m}\right).
\end{align}
Eq.\eqref{eq: coupling down superoperator} has also introduced the super-index $\bar{j}_{r} = \{\alpha_{j_{r}},\bar{\sigma}_{j_{r}},\ell_{j_{r}},m_{j_{r}}\}$.

Within the HEOM approach, one needs to converge the dynamics with respect to both the fermionic tier, which roughly corresponds to how many higher-order interactions with the lead are relevant to the transport, and with respect to the size of the vibrational basis. All results in this work were converged at a fermionic tier of either $2$ or $3$ and a vibrational basis size of $15$ to $25$.

\subsection{Reservoir Treatment of the Vibrational Degrees of Freedom} \label{subsec: Reservoir treatment of the vibrational degrees of freedom}

The HEOM formulation in Eq.\eqref{eq: HEOM Final} treats all degrees of freedom within the molecular Hamiltonian on an equal footing within $\rho(t)$. In regimes of vibrational instability, where the vibrational excitation is large, it can be more numerically efficient to treat the corresponding vibrational degrees of freedom as an extra reservoir. This idea has already been explored in Ref.\ \cite{Schinabeck2018,Mangaud2019,Liu2014,Duan2016,Dijkstra2017}, albeit under the assumption that \textit{all} vibrational degrees of freedom would be treated within the reservoir formulation. In scenarios where some vibrational degrees of freedom exhibit instability while some do not, perhaps due to different electronic-vibrational couplings or frequencies, it would be useful to have a framework that allows strongly-coupled vibrational modes to be kept within the reduced density matrix and weakly-coupled modes to receive the reservoir treatment.

This essentially amounts to rewriting the junction Hamiltonian as 
\begin{align}
H = & H_{\text{mol}} + H_{\text{res}} + H_{\text{mol-res}}, \label{eq: reservoir treatment total Hamiltonian} \\
H_{\text{mol}} = & \sum_{m} \varepsilon_{m} d^{\dag}_{m}d^{\ph}_{m} + \sum_{m < m'} U_{mm'}d^{\dag}_{m}d^{\ph}_{m}d^{\dag}_{m'}d^{\ph}_{m'} \nonumber \\
 & + \sum_{m}\sum_{i=1}^{N_{\text{sys}}}\lambda_{i}\left(b^{\ph}_{i} + b^{\dag}_{i}\right) d^{\dag}_{m}d^{\ph}_{m} + \sum_{i=1}^{N_{\text{sys}}} \omega^{\ph}_{i} b^{\dag}_{i}b^{\ph}_{i},  \label{eq: reservoir treatment molecular Hamiltonian} \\
H_{\text{res}} = & \sum_{\alpha,k\alpha} \varepsilon^{\ph}_{k\alpha} c^{\dag}_{k\alpha}c^{}_{k\alpha} + \sum_{k=1}^{N_{\text{res}}} \omega^{\ph}_{k} b^{\dag}_{k}b^{\ph}_{k}, \label{eq: reservoir treatment leads Hamiltonian} \\ 
H_{\text{mol-res}} = & \sum_{m}\sum_{\alpha,k\alpha}V_{m,\alpha}\left(c^{\dag}_{k\alpha}d^{\ph}_{m} + d^{\dag}_{m}c^{}_{k\alpha}\right) \nonumber \\
& + \sum_{m}\sum_{i=1}^{N_{\text{res}}}\lambda^{\ph}_{i}\left(b^{\ph}_{i} + b^{\dag}_{i}\right) d^{\dag}_{m}d^{\ph}_{m}, \label{eq: reservoir treatment molecular-lead Hamiltonian} 
\end{align}
where $N_{\text{sys}}$ and $N_{\text{res}}$ are the number of modes treated in the system and reservoir, respectively. Note that Eqs. \eqref{eq: reservoir treatment molecular Hamiltonian} and \eqref{eq: reservoir treatment leads Hamiltonian} exclude the zero point energies of the harmonic oscillators, which, since they just add a constant to the Hamiltonian, do not affect the dynamics.

%Note that the vibrational degrees of freedom in Eqs. \eqref{eq: reservoir treatment total Hamiltonian}-\eqref{eq: reservoir treatment molecular-lead Hamiltonian} have been written exclusively with bosonic creation and annihilation operators, which is required for the reservoir formulation of HEOM. 

Because the vibrational modes receiving the reservoir treatment are harmonic and the corresponding electronic-vibrational coupling is linear in $\hat{x}_{k}$, the effect of the $k$th vibrational mode in the reservoir on the molecule is also described by a two-time correlation function,
\begin{align}
C^{v}_{\text{vib},k} & = \langle b^{v}_{k}(t)b^{\bar{v}}_{k}(\tau)\rangle_{\text{vib,res}}
\end{align}
where $v = \pm$, $v = \mp$, $b^{-}_{k} = b^{\ph}_{k}$, and $b^{+}_{k} = b^{\dag}_{k}$. Application of the bath interaction picture now means that vibrational operators treated in the reservoir are time-dependent, $b^{v}_{k}(t) =  e^{i H_{\text{res}}t}b^{v}_{k}e^{-i H_{\text{res}}t} = e^{vi\omega_{k}t}b^{v}_{k}$. In order to treat these modes within the HEOM framework, the reservoir vibrational initial state must be chosen so that they are in thermal equilibrium with some temperature, $T_{\text{vib}}$,
\begin{align}
\rho_{\text{vib,res}}(0) & = \frac{e^{-H_{\text{vib,res}} /k_{B}T_{\text{vib}}}}{\text{Tr}_{\text{vib,res}}\left[e^{-H_{\text{vib,res}} /k_{B}T_{\text{vib}}}\right]},
\end{align}
and that the total initial state still factorizes as in Eq.\eqref{eq: factorized initial state}. Under these assumptions, the bath-correlation functions describing the vibrational reservoir decompose analytically into 
\begin{align}
C^{v}_{\text{vib},k} & = \eta_{k,v} e^{-\kappa_{k,v}t},
\end{align}
where $\kappa_{k,v} = \bar{v}i\omega_{k}$, $\eta_{k,+} = \bar{N}_{k}(0)$,  $\eta_{k,-} = 1 + \bar{N}_{k}(0)$, and $\bar{N}_{k}(0)$ is the initial excitation of mode $k$, $\bar{N}_{k}(0) = (1 + e^{\omega_{k}/k_{B}T_{\text{vib}}})^{-1}$. For steady-state observables, the specific choice of $T_{\text{vib}}$ is not particularly important, although it is highly relevant for transient dynamics.

The corresponding HEOM now has two hierarchies arising from the original fermionic electrodes and the new bosonic baths,
\begin{align}
\frac{\partial}{\partial t}\rho^{(n,q)}_{\bm{j}|\bm{q}} = & -i\left[H_{\text{mol}},\rho^{(n,q)}_{\bm{j}|\bm{q}} \right] \nonumber \\
& -\left(\sum_{r = 1}^{n} \kappa^{}_{j_{r}} + \sum_{k,v}\kappa_{k,v}q_{k,v}\right)\rho^{(n,q)}_{\bm{j}|\bm{q}} \nonumber \\
& - i\sum_{r=1}^{n} (-1)^{n-r}\mathcal{C}_{j_{r}}\rho^{(n-1),q}_{\bm{j}^{-}|\bm{q}} - i\sum_{j} \mathcal{A}^{\bar{\sigma}}_{\alpha,m
} \rho^{(n+1),q}_{\bm{j}^{+}| \bm{q}} \nonumber \\
&  -i\sum_{k,v} q_{k,v}\mathcal{C}^{\text{vib}}_{k,v} \rho^{(n,q-1)}_{\bm{j}|\bm{q}^{-}} - i \sum_{k} \mathcal{A}^{\text{vib}}_{k} \sum_{v} \rho^{(n,q+1)}_{\bm{j}|\bm{q}^{+}}, \label{eq: HEOM reservoir treatment of vibrations}
\end{align}
with the new vibrational superoperators
\begin{align}
\mathcal{A}^{\text{vib}}_{k} \rho^{(n,q)}_{\bm{j}|\bm{q}} & = \lambda_{k} \left( d^{\dag}_{}d \rho^{(n,q)}_{\bm{j}|\bm{q}} - \rho^{(n,q)}_{\bm{j}|\bm{q}}d^{\dag}_{}d \right), \\
\mathcal{C}^{\text{vib}}_{k} \rho^{(n,q)}_{\bm{j}|\bm{q}} & =  \lambda_{k} \left(\eta^{}_{k,v}d^{\dag}_{}d^{}_{}\rho^{(n,q)}_{\bm{j}|\bm{q}} - \eta^{*}_{k,\bar{v}}\rho^{(n,q)}_{\bm{j}|\bm{q}}d^{\dag}_{}d^{}_{}\right).
\end{align}
The ADOs of $q$th bosonic tier and $n$th fermionic tier now contain not just information about interactions with the metallic leads, but also $q$th-order interactions with vibrational modes in the reservoir. It is important to note that the truncation tier, $q$, and super-indices, $\bm{q}$, do not refer to specific operators. Rather, each $q$th-tier ADO is described by a vector of super-indices, $\bm{q} = (q^{-}_{0},q^{+}_{0},q^{-}_{1},\dots,q^{+}_{N_{\text{res}}})$, where each index $q^{v}_{k}$ is an integer whose value refers to the number of $b^{v}_{k}$ contained within this ADO. Alternatively, one can think of $q^{v}_{k}$ as the number of interactions of type $v$ with mode $k$ that this ADO describes. The bosonic tier, $q$, is then calculated by summing over all of these interactions: $q = \sum_{k} (q^{+}_{k} + q^{-}_{k})$. For a more detailed formulation, we refer to Ref.\ \cite{Schinabeck2018}, specifically Eq. (24).

This updated reservoir formulation of the HEOM was implemented via an extension of the solver introduced in Ref.\ \cite{Kaspar2021}, which uses an efficient iterative scheme to solve the HEOM directly for the steady state. Similar to the fermionic hierarchy, one needs to converge the new bosonic hierarchy with respect to the tier, $q$. All results in this work were converged with a bosonic tier between $10$ and $20$.

\subsection{Transport Observables from the Quantum HEOM Approach} \label{subsec: Transport observables from quantum theory}

The ADOs are not just mathematical objects required for the exact time-evolution of $\rho(t)$. Rather, they also provide direct information on bath properties, such as heat and charge flow. The average electric current through electrode $\alpha$, for example, can be directly obtained from ADOs that are $1$st-tier in the fermionic hierarchy \cite{Jin2008}. When all vibrational degrees of freedom are treated within the molecular density matrix, as in Sec.\ \ref{subsec: HEOM}, this yields 
\begin{align}
\langle I_{\alpha} \rangle(t) & = 2e \sum_{m,\ell}V_{m,\alpha}\text{Im}\left\{\text{Tr}_{\text{mol}}\left[d^{\ph}_{m}\rho^{(1)}_{K,+,\ell,m}(t)\right]\right\}.
\end{align}
In the reservoir treatment introduced in Eq.\eqref{eq: HEOM reservoir treatment of vibrations}, the current is $1$st-tier in the fermionic hierarchy and $0$th-tier in the bosonic hierarchy \cite{Schinabeck2018}, 
\begin{align}
\langle I_{\alpha} \rangle(t) & = 2e \sum_{m,\ell}V_{m,\alpha}\text{Im}\left\{\text{Tr}_{\text{mol}}\left[d^{\ph}_{m}\rho^{(1,0)}_{K,+,\ell,m}(t)\right] \right\}.
\end{align}
The steady-state average vibrational excitation of mode $i$, $\langle N_{i}\rangle(t) = \langle b^{\dag}_{i}b^{\ph}_{i}\rangle(t) $, on the other hand, is calculated in a very different manner between the two treatments. When mode $i$ is treated as part of the molecular Hamiltonian, the excitation is 
\begin{align}
\langle N_{i}\rangle(t) & = \text{Tr}_{\text{mol}}\left[b^{\dag}_{i}b^{\ph}_{i}\rho(t)\right].
\end{align}
When mode $i$ is treated as a bosonic reservoir, $b^{\dag}_{i}b^{\ph}_{i}$ is now an observable $2$nd order in bath operators, so it can be reconstructed from ADOs $2$nd-tier in the bosonic hierarchy \cite{Schinabeck2018},
\begin{align}
\langle N_{i} \rangle(t) & = \bar{N}_{i}(0) + \text{Tr}_{\text{mol}}\left\{\rho^{(0,2)}_{| 1_{i,+}1_{i,-}}(t)\right\},
\end{align}
where $ \bar{N}_{i}(0)$ is the initial excitation due to choosing a thermalized initial state. 

\section{Electronic Friction and Langevin Dynamics: The HEOM-LD Approach} \label{sec: Electronic friction and Langevin dynamics}

In Ref.\ \cite{Rudge2023}, two of the authors formulated a mixed quantum-classical version of the HEOM, which, for self-completeness, is outlined here. From it, one can calculate quantum electronic forces acting on classical vibrational degrees of freedom. This section also describes how one uses these electronic forces in the corresponding Langevin dynamics to form the complete HEOM-LD method.

\subsection{Derivation of the HEOM-LD approach} \label{subsec: Deriving the HEOM-LD approach}

Often, vibrational modes in the molecule can be separated into $N_{\text{qu}}$ high-frequency modes that need to be treated quantum mechanically, $(\hat{\boldsymbol{x}}_{\text{qu}},\hat{\boldsymbol{p}}_{\text{qu}})$, and $N_{\text{cl}}$ low-frequency modes that can be well-approximated with a classical treatment, $(\boldsymbol{x}_{\text{cl}},\boldsymbol{p}_{\text{cl}})$. As discussed in Ref.\ \cite{Rudge2023}, this amounts to Wigner transforming the HEOM with respect to the low-frequency vibrational degrees of freedom and truncating the resulting expansion for terms linear in $\hbar$ and higher. In the joint Liouville space of all ADOs, this yields
\begin{align}
\dot{\boldsymbol{\tilde{\rho}}}(\bm{x}_{\text{cl}},\bm{p}_{\text{cl}} ; t) & = \tilde{\mathcal{L}}(\bm{x}_{\text{cl}})\boldsymbol{\tilde{\rho}}(\bm{x}_{\text{cl}},\bm{p}_{\text{cl}} ; t) \nonumber \\
& \:\:\:\:\:\: + \{\!\!\{\tilde{H}_{\text{mol}}(\bm{x}_{\text{cl}},\bm{p}_{\text{cl}}),\boldsymbol{\tilde{\rho}}(\bm{x}_{\text{cl}},\bm{p}_{\text{cl}} ; t) \}\!\!\}_{a}, \label{eq: QCLE final}
\end{align}
where $\tilde{O}$ is the Wigner transform of operator $\hat{O}$ and $\boldsymbol{\tilde{\rho}} = \left[\tilde{\rho},\tilde{\rho}^{(1)}_{j_{1}},\dots,\tilde{\rho}^{(N_{\text{max}})}_{\bm{j}}\right]^{T}$. The leading-order contribution in the expansion of the Wigner-transformed HEOM corresponds to the first term in Eq.\eqref{eq: QCLE final}, which is proportional to $1/\hbar$ and contains the HEOM dynamics frozen at one classical vibrational frame, $\tilde{\mathcal{L}}(\bm{x}_{\text{cl}})$. The equation of motion for the $n$th-tier ADO described by this superoperator is 
\begin{align}
\dot{\tilde{\sigma}}^{(n)}_{\bm{j}}(\bm{x}_{\text{cl}},t) & = -i\left[\tilde{H}_{\text{mol}}(\bm{x}_{\text{cl}}),\tilde{\sigma}^{(n)}_{\bm{j}}\right]  - \left(\sum_{r = 1}^{N} \kappa^{}_{j_{r}}\right)\tilde{\sigma}^{(n)}_{\bm{j}} \nonumber \\
& \:\:\:\:\:  - i\sum_{r=1}^{n} (-1)^{n-r}\mathcal{C}_{j_{r}}\tilde{\sigma}^{(n-1)}_{\bm{j}^{-}} - i\sum_{j} \mathcal{A}^{\bar{\sigma}}_{\alpha,m} \tilde{\sigma}^{(n+1)}_{\bm{j}^{+}}. \label{eq: HEOM frozen}
\end{align}
Here, the ADOs are written with $\sigma$ and not $\rho$ so as to distinguish between the objects following the time-evolution in Eq.\eqref{eq: HEOM frozen} and the objects following the time-evolution in Eq.\eqref{eq: QCLE final}.

The second term in Eq.\eqref{eq: QCLE final}, meanwhile, arises from the next-to-leading-order term in the expansion of the Wigner-transformed HEOM, which is $0$th-order in $\hbar$. Similar to a quantum-classical Liouville equation, this contains a symmetrized Poisson bracket with each ADO \cite{Kapral1999,Dou2016a},
\begin{align}
& \{\!\!\{\tilde{H}_{\text{mol}}(\bm{x},\bm{p}),\boldsymbol{\tilde{\rho}}(\bm{x},\bm{p} ; t) \}\!\!\}_{a} = \nonumber \\
& \:\:\:\:\: \left[\{\tilde{H}_{\text{mol}},\tilde{\rho} \}_{a},\{\tilde{H}_{\text{mol}},\tilde{\rho}^{(1)}_{j_{1}} \}_{a},\dots,\{\tilde{H}_{\text{mol}},\tilde{\rho}^{(N_{\text{max}})}_{\bm{j}}\}_{a}\right]^{T},
\end{align}
where $\{A_{1},A_{2}\}_{a} = \frac{1}{2}\left(\{A_{1},A_{2}\} - \{A_{2},A_{1}\}\right)$ and
\begin{align}
\{A_{1},A_{2}\} & = \sum_{i = 1}^{N_{\text{cl}}} \left(\frac{\partial A_{1}}{\partial x_{i}}\frac{\partial A_{2}}{\partial p_{i}} - \frac{\partial A_{1}}{\partial p_{i}}\frac{\partial A_{2}}{\partial x_{i}}\right).
\end{align}

A Langevin equation of motion for the classical vibrational modes can then be derived by directly applying a timescale separation between quantum and classical degrees of freedom to the equation of motion for $A(\bm{x}_{\text{cl}},\bm{p}_{\text{cl}} ; t) = \text{Tr}_{\text{mol,qu}}\left[\tilde{\rho}(\bm{x}_{\text{cl}},\bm{p}_{\text{cl}} ; t)\right]$, which is the phase-space probability density of the classical vibrational modes. Note that the $\text{Tr}_{\text{mol,qu}}\left[\dots\right]$ is now a trace over the quantum degrees of freedom remaining in the molecular Hamiltonian after the Wigner transform and that these degrees of freedom may include any high-frequency vibrational modes requiring a quantum mechanical treatment. One expects this to be justified if the timescale of electronic relaxation is much faster than the timescale of the vibrational dynamics. In scenarios where the molecule-lead coupling is larger than the effective vibrational frequency, $\Gamma,k_{B}T \gg \omega$, such as the dynamics of heavy molecules, the HEOM-LD should perform well. At high temperatures, furthermore, not only are quantization effects negligible, which justifies a classical treatment of the vibrations, but the Markovian approximation of electronic dynamics is also valid. 

After these two approximations, the classical vibrational modes in the molecule follow a Markovian Langevin equation of motion, which for the $i$th coordinate is 
\begin{align}
m_{i}\ddot{x}_{\text{cl},i} & =  F^{\text{ad}}_{i}(\bm{x}_{\text{cl}}) - \sum_{j} \gamma_{ij}(\bm{x}_{\text{cl}}) \dot{x}_{\text{cl},j} + f_{i}(t), \label{eq: Langevin Markovian final}
\end{align}
where $f_{i}(t)$ is a Gaussian random force with white noise,
\begin{align}
\langle f_{i}(t)f_{j}(t')\rangle & = 2 D_{ij}(\bm{x}_{\text{cl}}) \delta(t - t'). \label{eq: correlation function Markovian}
\end{align}
Here, $ D_{ij}(\bm{x}_{\text{cl}})$ is the strength of the correlation function of the stochastic force between vibrational coordinates $i$ and $j$. 

The adiabatic contribution to the mean force, $F^{\text{ad}}_{i}(\bm{x}_{\text{cl}})$, as well as the force difference operator, $\delta \hat{F}^{\text{ad}}_{i}(\bm{x}_{\text{cl}})$, are
\begin{align}
F^{\text{ad}}_{i}(\bm{x}_{\text{cl}}) & = - \text{Tr}_{\text{mol,qu}}\left[\partial_{i} \tilde{H}_{\text{mol}}(\bm{x}_{\text{cl}}) \boldsymbol{\tilde{\sigma}}^{\text{ss}}(\bm{x}_{\text{cl}})\right], \label{eq: average electronic force HEOM 1} \\ 
\delta \hat{F}_{i}(\bm{x}_{\text{cl}}) & = \partial_{i}\tilde{H}_{\text{mol}}(\bm{x}_{\text{cl}}) - F^{\text{ad}}_{i}(\bm{x}_{\text{cl}}). \label{eq: force difference operator}
\end{align}
Here, $\partial_{i}$ is shorthand for a derivative with respect to the $i$th classical vibrational coordinate, $\partial_{i} = \partial/\partial x_{\text{cl},i}$, and $\boldsymbol{\tilde{\sigma}}^{\text{ss}}(\bm{x}_{\text{cl}})$ is the steady-state joint density operator of the HEOM frozen at one classical vibrational frame, 
\begin{align}
\tilde{\mathcal{L}}(\bm{x}_{\text{cl}})\boldsymbol{\tilde{\sigma}}^{\text{ss}}(\bm{x}_{\text{cl}}) & = \bm{0}. \label{eq: adiabatic HEOM}
\end{align}

Eq.\eqref{eq: Langevin Markovian final} also introduces the electronic friction tensor as well as the correlation function of the stochastic force,
\begin{align}
\gamma_{ij}(\bm{x}_{\text{cl}}) & = -\lim_{\eta \rightarrow 0^{+}} \int^{\infty}_{0} dt \: \text{Tr}_{\text{mol,qu}}\left[\partial_{i} \tilde{H}_{\text{mol}}e^{-(\tilde{\mathcal{L}} + \eta)t}\partial_{j} \boldsymbol{\tilde{\sigma}}^{\text{ss}}\right], \label{eq: friction calculation} \\
D_{ij}(\bm{x}_{\text{cl}}) & = \frac{1}{2}\lim_{\eta \rightarrow 0^{+}} \int^{\infty}_{0} dt \: \text{Tr}_{\text{mol,qu}}\Big[\delta \hat{F}_{i}e^{-(\tilde{\mathcal{L}} + \eta)t}\times \nonumber \\ 
& \hspace{4cm} \left(\delta \hat{F}_{j}\boldsymbol{\tilde{\sigma}}^{\text{ss}} + \boldsymbol{\tilde{\sigma}}^{\text{ss}}\delta \hat{F}_{j}\right)\Big].\label{eq: correlation function of the stochastic force calculation HEOM}
\end{align}
At equilibrium, the friction tensor is always symmetric and positive-definite, such that it has an overall damping effect on the classical vibrational motion \cite{Dou2017c,Rudge2023}. It is also related to the correlation function of the stochastic force via the fluctuation-dissipation relation,
\begin{align}
D_{ij}(\bm{x}_{\text{cl}}) & = \gamma_{ij}(\bm{x}_{\text{cl}}) k_{B}T 
\end{align}
Out of equilibrium, however, the fluctuation-dissipation theorem and the positive-definiteness of the friction tensor are not guaranteed. At finite bias voltages, for example, molecular vibrations still have excitation and relaxation mechanism from coupling to electron-hole pairs near the chemical potentials of the leads, but are also excited by the nonequilibrium electric current, which manifests as 
\begin{align}
D_{ij}(\bm{x}_{\text{cl}}) \geq \gamma_{ij}(\bm{x}_{\text{cl}}) k_{B}T.
\end{align} 
Although the theory here focuses on electronic friction, the calculation of friction and diffusion tensors is also well established for classical molecular simulations \cite{Straub1987,Hummer2005,Hegger2009,Wolf2018}.

To end this subsection, note that one could also solve the corresponding Fokker-Planck equation for $A(\bm{x}_{\text{cl}},\bm{p}_{\text{cl}} ; t)$,
\begin{align}
\frac{\partial A}{\partial t} & = - \sum_{i}\frac{p_{\text{cl},i}}{m_{i}} \frac{\partial A}{\partial x_{\text{cl},i}} + \sum_{ij}\gamma_{ij}(\mathbf{x}_{\text{cl},i}) \frac{\partial}{\partial p_{\text{cl},i}}\left(\frac{p_{\text{cl},j}}{m_{j}}A\right) \nonumber \\
& \:\:\:\:\:\: + \sum_{i}F^{}_{i}(\mathbf{x}_{\text{cl}})\frac{\partial A}{\partial p_{\text{cl},i}}   + \sum_{ij}D_{ij}(\mathbf{x}_{\text{cl}}) \frac{\partial^{2}A}{\partial p_{\text{cl},i}\partial p_{\text{cl},j}}. \label{eq: FP final}
\end{align}

\begin{center}
\begin{figure*}
\begin{subfigure}[b]{0.0\textwidth}
\phantomcaption\label{fig: 1a}
\end{subfigure}
\begin{subfigure}[b]{0.0\textwidth}
\phantomcaption\label{fig: 1b}
\end{subfigure}
\begin{subfigure}[b]{0.0\textwidth}
\phantomcaption\label{fig: 1c}
\end{subfigure}
\begin{subfigure}[b]{0.0\textwidth}
\phantomcaption\label{fig: 1d}
\end{subfigure}
\begin{subfigure}[b]{0.0\textwidth}
\phantomcaption\label{fig: 1e}
\end{subfigure}
\begin{subfigure}[b]{0.0\textwidth}
\phantomcaption\label{fig: 1f}
\end{subfigure}
\begin{centering}
\includegraphics[width = 0.8\textwidth]{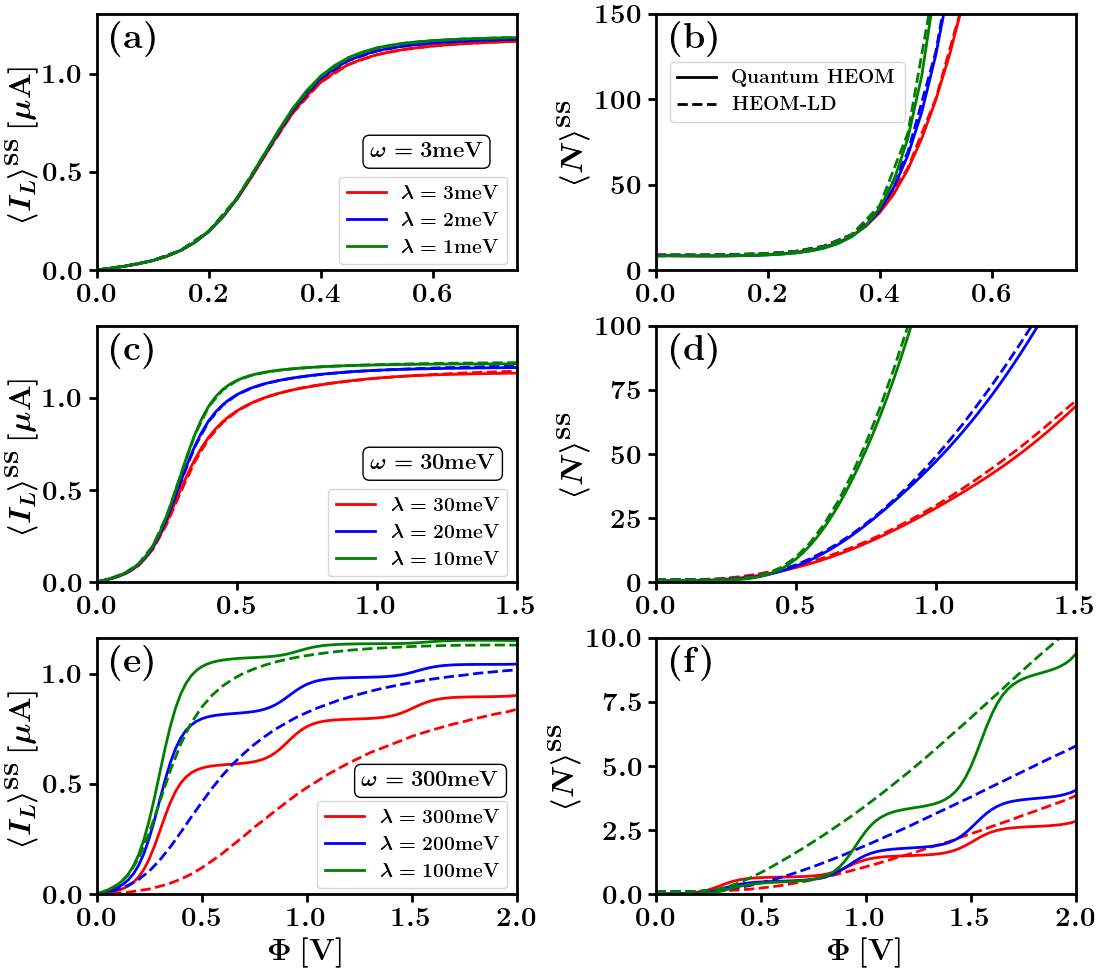}
\caption{Steady-state transport observables as a function of bias voltage for the 1L1M model. Left column: Steady-state electronic current, $\langle I_{L} \rangle^{\text{ss}}$. Right column: Corresponding steady-state vibrational excitation, $\langle N \rangle^{\text{ss}}$. The top row shows results for a vibrational frequency of $\omega = 3\text{ meV}$, the middle a vibrational frequency of $\omega = 30\text{ meV}$, and the bottom a vibrational frequency of $\omega = 300\text{ meV}$. For each vibrational frequency, results for a range of electronic-vibrational couplings are also plotted, which are indicated by color in the plots. Other parameters are $\Gamma = 20\text{ meV}$ and $k_{B}T = 25.8\text{ meV}$. The electronic energy is chosen such that the energy level after the small polaron shift is always $\varepsilon_{0} = \tilde{\varepsilon}_{0} - \frac{\lambda^{2}}{\omega} = 150\text{ meV}$.}
\label{fig: 1}
\end{centering}
\end{figure*}
\end{center}

\subsection{Solving the Langevin equation}

Due to the presence of the stochastic force, propagating the classical dynamics accurately for long times can be challenging. Modern techniques for solving Langevin equations generally use integrators based on a Trotter decomposition of the propagator defined by the Fokker-Planck equation \cite{Kieninger2023}. Restricting Eq.\eqref{eq: FP final} to one dimension for notational simplicity, the time-evolution operator of the classical phase-space probability density, $\mathcal{L}_{\text{cl}}$, can be split as 
\begin{align}
\mathcal{L}_{\text{cl}} = & \underbrace{-\frac{p_{\text{cl}}}{m} \frac{\partial }{\partial x_{\text{cl}}}}_{\mathcal{L}_{A}} + \underbrace{F\frac{\partial}{\partial p_{\text{cl}}}}_{\mathcal{L}_{B}} + \underbrace{\gamma \frac{\partial}{\partial p_{\text{cl}}}\frac{p_{\text{cl}}}{m} + D \frac{\partial^{2}}{\partial p_{\text{cl}}^{2}}}_{\mathcal{L}_{O}}.
\end{align}
The classical time-evolution is defined by the propagator $e^{-\Delta t \mathcal{L}_{\text{cl}}}$, but one can construct a family of approximations via the Trotter decomposition \cite{Bussi2007}, 
\begin{align}
e^{-\Delta t\mathcal{L}_{\text{cl}}} \approx & \prod^{1}_{j = 3}e^{-(\Delta t/2)\mathcal{L}_{j}} \prod^{3}_{k = 1}e^{-(\Delta t/2)\mathcal{L}_{k}}, \label{eq: Trotter decomposition}
\end{align}
where $j \in \{A,B,O\}$. Since the propagators $e^{-(\Delta t/2)\mathcal{L}_{j}}$ do not commute, different orderings within Eq.\eqref{eq: Trotter decomposition} yield different approximations. In generating the results in Sec.\ \ref{sec: Results}, both the ABOBA \cite{Sachs2017} and OBABO \cite{Bussi2007} algorithms were tested, where the letters denote the order within Eq.\eqref{eq: Trotter decomposition}. Drawing from our experience with numerical integration of Markovian Langevin equations, we know that, for large-dimensional systems, the OBABO algorithm has been shown to be generally more accurate because it splits the stochastic part of the propagation \cite{Wolf2020,Kieninger2023}. In all the one-dimensional examples discussed in Sec.\ \ref{sec: Results}, however, there was no discernible difference in accuracy or convergence speed between the two algorithms.

\subsection{Transport Observables from the HEOM-LD Approach}

Consider a general operator treated within the quantum part of the system in the HEOM-LD approach, $O$, which does not explicitly depend on classical coordinates. At a particular phase-space coordinate, $(\bm{x}_{\text{cl}},\bm{p}_{\text{cl}})$, it will have an instantaneous average,
\begin{align}
\langle O \rangle^{\ph}_{\text{qu}}(\bm{x}_{\text{cl}},\bm{p}_{\text{cl}} ; t ) & = \text{Tr}_{\text{qu}} \left\{O \tilde{\rho}^{\ph}_{\text{total}}(\bm{x}_{\text{cl}},\bm{p}_{\text{cl}} ; t )\right\}. \label{eq: instantaneous full average}
\end{align}
Here, $\tilde{\rho}_{\text{total}}$ is the Wigner transform of the total density matrix with respect to the classical vibrational degrees of freedom and the subscript $\text{qu}$ indicates an average over only the quantum part of the system. One calculates the actual expectation value, $\langle O \rangle(t)$, by averaging $\langle O \rangle^{\ph}_{\text{qu}}$ over the classical part of the system,
\begin{align}
\langle O \rangle (t) & = \int d\bm{x}_{\text{cl}}d\bm{p}_{\text{cl}} \: A(\bm{x}_{\text{cl}},\bm{p}_{\text{cl}} ; t) \langle O \rangle_{\text{qu}}^{\text{ss}}(\bm{x}_{\text{cl}},\bm{p}_{\text{cl}}). \label{eq: full average}
\end{align}
In this work, the instantaneous quantum averages will be approximated adiabatically, such that $\tilde{\rho}^{\ph}_{\text{total}}(\bm{x}_{\text{cl}},\bm{p}_{\text{cl}} ; t )$ is replaced by $\sigma^{\text{ss}}(\bm{x}_{\text{cl}})$, the adiabatic steady-state density matrix frozen at vibrational coordinate $\bm{x}_{\text{cl}}$: see Eq.\eqref{eq: adiabatic HEOM}. For example, the instantaneous average current and electronic occupations are approximated as 
\begin{gather}
\langle I_{\alpha}\rangle_{\text{qu}} \approx 2e \sum_{\ell,m}V_{\alpha,m}\text{Im}\left\{\text{Tr}_{\text{mol,qu}}\left[d^{\ph}_{m}\sigma^{(1),\text{ss}}_{K,+,\ell,m}(\bm{x}_{\text{cl}})\right]\right\}, \\
\langle d^{\dag}_{m}d^{\ph}_{m} \rangle_{\text{qu}} \approx \text{Tr}_{\text{mol,qu}}\left[d^{\dag}_{m}d^{\ph}_{m}\sigma^{\text{ss}}(\bm{x}_{\text{cl}})\right],
\end{gather}
which is now only a function of $\mathbf{x}_{\text{cl}}$. If the quantum part of the system contains vibrational mode $i$, then this same approach would yield the following approximation for the instantaneous vibrational excitation,
\begin{align}
\langle N_{i} \rangle_{\text{qu}}(t) \approx & \text{Tr}_{\text{mol,qu}}\left[b^{\dag}_{i}b^{\ph}_{i}\sigma^{\text{ss}}(\bm{x}_{\text{cl}})\right].
\end{align}

In the following section, only steady-state quantities will be discussed. These will be denoted with an ss superscript and can be obtained from the above approach by taking the long-time limit of the phase-space probability density. Since in this work Langevin and not Fokker-Planck dynamics are used, the classical averages are calculated by averaging over many trajectories, rather than directly evaluating Eq.\eqref{eq: full average}. To generate the trajectories and sample points, the following algorithm was used. 

First, the electronic forces and instantaneous steady-state expectation values of quantum operators are calculated via the procedure outlined in Sec.\ \ref{subsec: Deriving the HEOM-LD approach}. A large number of Langevin trajectories, $N_{\text{traj}}$, are initialized. The initial coordinates and momenta for each trajectory are sampled from the Wigner function of the initial quantum state, which provides the initial distribution for the classical calculation \cite{Erpenbeck2018}. Since all systems considered in this work contain harmonic oscillators and are one-dimensional in the classical coordinates, this is always chosen to be the ground state of the harmonic oscillator of the uncharged electronic state, which in dimensionless coordinates is
\begin{align}
\rho_{W}(x_{\text{cl}},p_{\text{cl}} ; 0) = & \frac{1}{\pi} e^{-(x_{\text{cl}}^{2} + p_{\text{cl}}^{2})}. 
\end{align}
Because the Fokker-Planck equation for the systems treated here has a unique steady-state, the exact choice of $\rho_{W}(x_{\text{cl}},p_{\text{cl}} ; 0)$ is, in this work, relatively unimportant. However, if one were interested in transient dynamics, then the choice of $\rho(0)$ would be crucial. 

Each trajectory is propagated to a long-enough time, $t_{\text{ss}}$, such that $A(\bm{x}_{\text{cl}},\bm{p}_{\text{cl}} ; t_{\text{ss}}) \rightarrow A^{\text{ss}}(\bm{x}_{\text{cl}},\bm{p}_{\text{cl}})$. In practice, this is converged by first running a low number of trajectories and observing the time-evolution of the average kinetic and potential energies. Once a trajectory reaches $t_{\text{ss}}$, it is propagated further and sampled $N_{\text{sample}}$ times, with some interval time, $t_{\text{int}}$, between each sample point to ensure independence. 

Finally, these $N_{\text{total}} = N_{\text{traj}}\times N_{\text{sample}}$ points are used to calculate the steady-state expectation values. For quantum operators, this is 
\begin{align}
\langle O \rangle^{\text{ss}} = & \left(\sum_{k = 1}^{N_{\text{total}}}\langle O \rangle^{\text{ss}}_{\text{qu}}(\bm{x}_{\text{cl},k}^{\text{ss}})\right)\Big/N_{\text{total}}.
\end{align}
Expectation values of classical observables are calculated in the same manner. Of particular interest is the steady-state excitation of the $i$th classical vibrational mode. For a harmonic oscillator, this can be approximated from $E_{i}$, the total energy contained in the mode:
\begin{align}
\langle N_{i} \rangle^{\text{ss}} \approx & \: E_{i}/\omega_{i} = \frac{1}{2}\left(\langle x_{i}^{2}\rangle^{\text{ss}} + \langle p_{i}^{2}\rangle^{\text{ss}}\right),
\end{align}
 In the quantum case, if one were to define the vibrational excitation via the energy, then there would be a contribution from the zero point energy, which is something that the classical approach cannot reproduce. However, as will become evident in the next section, the contribution from the zero point energy is generally small when the vibrational frequency is lower than the temperature and molecule-lead coupling: that is, in the parameter regimes for which HEOM-LD performs well. Finally, note that all results presented in Sec.\ \ref{sec: Results} have been calculated with $N_{\text{traj}} = N_{\text{sample}} = 500$. 

\section{Results}\label{sec: Results}

In this section, the steady-state transport properties of three reduced models for molecules in molecular junctions are explored. The first system, discussed in Sec.\ \ref{subsec: Assessing the adiabatic approximation via the Holstein model}, contains a single electronic level linearly coupled to a harmonic oscillator, which will be named the one-level, one-mode (1L1M) model and will be used to explore appropriate parameter regimes in the HEOM-LD approach. The second system extends this simple model by considering two electronic levels that are linearly coupled to a harmonic oscillator and have a Coulomb repulsion between them. This will be named the two-level, one-mode (2L1M) model and is analyzed in Sec.\ \ref{subsec: Adding an Electron-Electron interaction}. In Sec.\ \ref{subsec: One electronic level coupled to two vibrational modes with different timescales}, a single electronic level is linearly coupled to low-frequency oscillator, which is treated classically, and a high-frequency harmonic oscillator, which is treated quantum mechanically within the HEOM. 

%Again, the presence of a strong quantum interaction stabilizes the junction, as the two vibrational modes compete for the same electronic energy.

%The presence of the electron-electron interaction has a twofold effect. First, the accuracy of the HEOM-LD approach now depends strongly on the molecule-lead coupling, such that the timescale separation between fast electronic and slow vibrational degrees of freedom must be satisfied. Second, the interaction has a stabilizing effect on the nanojunction, which manifests through negative differential vibrational excitation as a function of bias voltage. 

\subsection{Basic Vibronic Model}\label{subsec: Assessing the adiabatic approximation via the Holstein model}

In the mass- and frequency-transformed coordinates, the molecular Hamiltonian is
\begin{align}
H_{\text{mol}} & = \left(\tilde{\varepsilon}_{0} + \lambda\sqrt{2}\:\hat{x}\right) d^{\dag}_{}d^{\ph} + \frac{\omega}{2}\left(\hat{p}^{2} + \hat{x}^{2}\right). \label{eq: basic vibronic Hamiltonian}
\end{align}
Given that there is only one electronic and vibrational degree of freedom and there are thus only a few parameters, it is an excellent testbed in which to assess the quality of the HEOM-LD approach. This is shown in Fig.\ (\ref{fig: 1}), where the steady-state electric current and vibrational excitation are plotted as a function of bias voltage for a range of vibrational parameters. Before starting the analysis, note that in Fig.\ (\ref{fig: 1}), the electronic energy level has been chosen such that the energy level after the small polaron shift is always $\varepsilon_{0} = \tilde{\varepsilon}_{0} - \frac{\lambda^{2}}{\omega} = 150 \text{ meV}$. This ensures that the onset of resonant transport occurs at the same voltage for all parameter regimes \cite{Lang1963}.

Fig.\ (\ref{fig: 1a}) and Fig.\ (\ref{fig: 1b}) depict transport observables in the adiabatic limit, where $\omega = 3\text{ meV}$, $\Gamma = 20 \text{ meV}$, and $k_{B}T = 25.8 \text{ meV}$. The electronic current displays a single step at the opening of the elastic channel, when $\varepsilon$ enters the bias window. Since $k_{B}T \gg \omega$, the vibrational mode behaves essentially classically and its excitation increases monotonically. Due to the small frequency and electronic-vibrational couplings, the mode experiences a vibrational instability at relatively low voltages, which arises from the well-known effect of current-induced heating \cite{Schinabeck2018}.  Since $\Gamma > \omega$, the timescale of vibrational motion is too slow for electrons to see anything but fixed nuclei during relaxation and the HEOM-LD approach exactly reproduces the steady-state observables calculated from the full quantum approach for all electronic-vibrational couplings. 

In the intermediate regime, shown in Fig.\ (\ref{fig: 1c}) and Fig.\ (\ref{fig: 1d}), the vibrational frequency is on the same order of magnitude as the temperature and molecule-lead coupling: $\omega = 30\text{ meV}$. Although the timescale separation is formally violated, the HEOM-LD approach still performs quite well. This can be attributed to two factors. First, since $k_{B}T \sim \omega$, the vibrational dynamics are still essentially classical, such that there is again only one step in the current at $e\Phi/2 = \varepsilon_{0}$. Second, whether a vibrational mode experiences adiabatic or nonadiabatic dynamics depends not just on the relative timescales of electronic and vibrational motion, but also strongly on the electronic-vibrational coupling. In Holstein-type models, this manifests via the separation between the ground and excited state diabats: the small polaron shift. Even in this intermediate regime, the largest shift is $\lambda^{2}/\omega = 30 \text{ meV} \sim \Gamma$, which ensures essentially adiabatic dynamics \cite{Eidelstein2013}. Finally, note that the mode also experiences a vibrational instability in this regime, although this occurs at higher voltages than in the classical regime. 

In the nonadiabatic regime, where $\Gamma, k_{B}T < \omega$, quantization of the vibrational energy becomes relevant to transport processes. Since the electronic transport occurs on a timescale slower than vibrational motion, the vibrational motion has sufficient time to respond during electronic relaxation, such that electrons occupying the molecule form polarons with vibrational phonons. Under a small polaron transformation, it can be shown that this rescales the molecule-lead coupling to a much smaller effective value, $\Gamma_{\text{eff}} \sim \Gamma e^{-(\lambda/\omega)^{2}}$ \cite{Lang1963}. As shown in Figs.\ (\ref{fig: 1e}) and (\ref{fig: 1f}), where $\omega = 300\text{ meV}$, this translates to additional steps in the average current and vibrational excitation, corresponding to the opening of phonon-assisted transport channels entering the bias window, $\varepsilon + \nu \omega$ with $\nu = 0,1,2,\dots$ \cite{Haertle2011a,Haertle2011b}. For $\lambda/\omega = 1$, the current and vibrational excitation are suppressed at low bias voltages due to the Franck-Condon blockade \cite{Koch2005,Koch2006a,Koch2006b}. While the mixed quantum-classical approach roughly predicts the quantitative behavior of the electronic current, especially at high bias voltages, it fails to reproduce the extra steps, which are a purely quantum effect. 

%Although the vibrational excitation in the quantum case is frozen out at low bias voltages \cite{Koch2005}, in the classical treatment, electrons can exchange any amount of energy with the vibrational mode, which means the HEOM-LD approach constantly overpredicts the vibrational excitation and, consequently, underpredicts the electric current.

In comparison to the parameter regimes explored above, many semiclassical treatments restrict $\omega$ to the meV range to ensure that the molecule-lead coupling or the temperature is much larger than the effective frequency \cite{Dou2018b,Dou2015b,Dou2015c,Dou2015d,Bode2011,Lue2012}. Considering that vibrational modes of actual molecules often have effective frequencies in the range of $0.01-0.4 \text{eV}$ \cite{Teale2004} and that realistic molecule-lead couplings also fall within this range, the results of this section demonstrate that the HEOM-LD approach can potentially be used in a wider parameter regime than what has been previously reported, at least for models in which the quantum part of the system is noninteracting.

\begin{figure*}
\begin{subfigure}[b]{0.0\textwidth}
\phantomcaption\label{fig: 2a}
\end{subfigure}
\begin{subfigure}[b]{0.0\textwidth}
\phantomcaption\label{fig: 2b}
\end{subfigure}
\begin{subfigure}[b]{0.0\textwidth}
\phantomcaption\label{fig: 2c}
\end{subfigure}
\begin{subfigure}[b]{0.0\textwidth}
\phantomcaption\label{fig: 2d}
\end{subfigure}
\begin{centering}
\includegraphics[width = 0.8\textwidth]{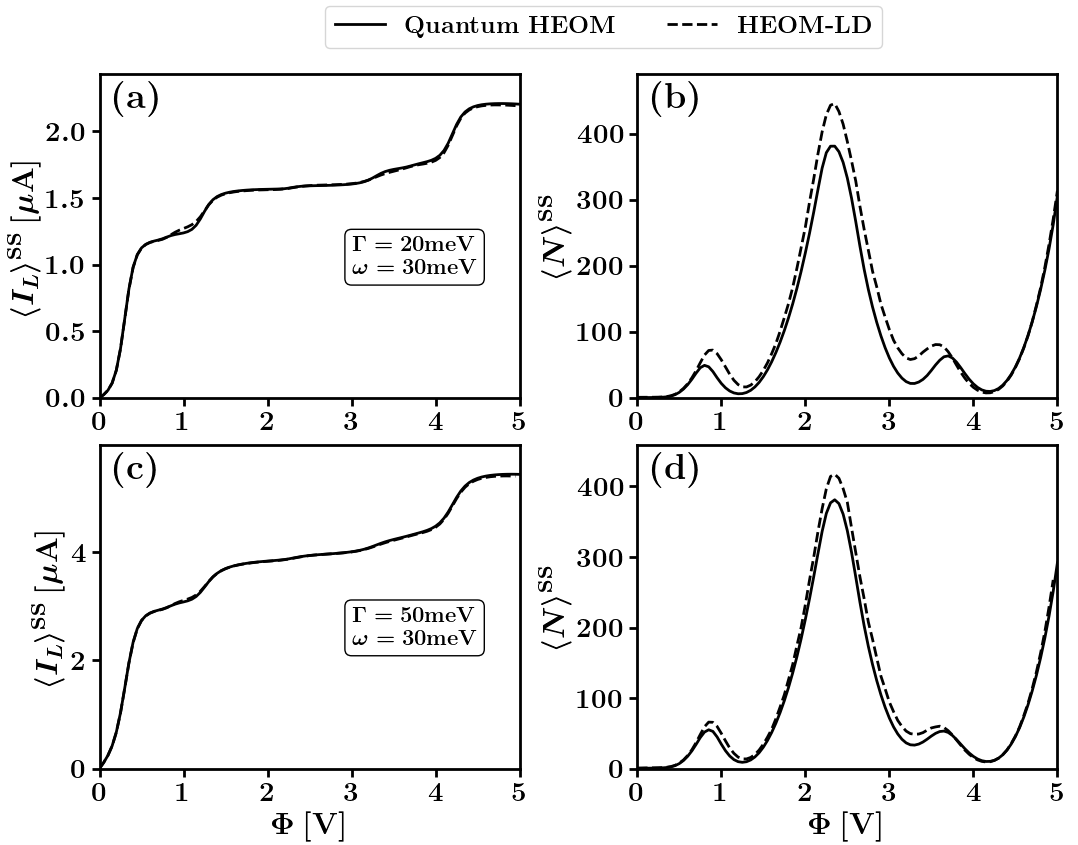}
\caption{Steady-state transport observables as a function of bias voltage for the two-level, one-mode model. Left column: Steady-state electronic current, $\langle I_{L} \rangle^{ss}$. Right column: Corresponding steady-state vibrational excitation, $\langle N \rangle^{ss}$. Both rows set $k_{B}T = 25.8\text{ meV}$, $\omega = 30\text{ meV}$, and $\lambda = 10\text{ meV}$, but change the molecule-lead coupling: in the top row, $\Gamma = 20\text{ meV}$, and in the bottom, $\Gamma = 50\text{ meV}$.  Electronic parameters within the molecule are chosen such that the energies after the small polaron shift are $\varepsilon^{\ph}_{1} = \tilde{\varepsilon}^{\ph}_{1} - \frac{\lambda^{2}}{\omega} = 150\text{ meV}$, $\varepsilon^{\ph}_{2} = \tilde{\varepsilon}^{\ph}_{2} - \frac{\lambda^{2}}{\omega} = 600\text{ meV}$, and $U = \tilde{U} - 2\frac{\lambda^{2}}{\omega} = 1.5\text{ eV}$.}
\label{fig: 2}
\end{centering}
\end{figure*}

\subsection{Influence of Electron-Electron Interaction} \label{subsec: Adding an Electron-Electron interaction}

In this subsection, the simple vibronic model is extended by adding an extra electronic level and an electron-electron interaction, forming the two-level, one-mode (2L1M) model, with Hamiltonian
\begin{align}
H_{\text{mol}} = & \tilde{\varepsilon}^{\ph}_{1}d^{\dag}_{1}d^{\ph}_{1} + \tilde{\varepsilon}^{\ph}_{2}d^{\dag}_{2}d^{\ph}_{2} + \tilde{U}d^{\dag}_{1}d^{\ph}_{1}d^{\dag}_{2}d^{\ph}_{2} \nonumber \\
& +\lambda\sqrt{2}\hat{x}\left(d^{\dag}_{1}d^{\ph}_{1} + d^{\dag}_{2}d^{\ph}_{2}\right) + \frac{\omega}{2}\left(\hat{p}^{2} + \hat{x}^{2}\right). \label{eq: general system Hamiltonian 2L1M}
\end{align}
In all calculations, both energy levels are coupled equally to both leads: $\Gamma_{\alpha,m} = \Gamma/2$. In addition, parameters are chosen such that the energies after the small polaron shift are $\varepsilon^{\ph}_{1} = \tilde{\varepsilon}^{\ph}_{1} - \frac{\lambda^{2}}{\omega} = 150\text{ meV}$, $\varepsilon^{\ph}_{2} = \tilde{\varepsilon}^{\ph}_{2} - \frac{\lambda^{2}}{\omega} = 600\text{ meV}$, and $U = \tilde{U} - 2\frac{\lambda^{2}}{\omega} = 1.5\text{ eV}$. 

The inclusion of the extra electronic level and electron-electron interaction has several effects on the steady-state transport properties, which are shown in Fig.\ (\ref{fig: 2}). First, the electric current now displays four steps at the voltages where the $\varepsilon^{\ph}_{1}$, $\varepsilon^{\ph}_{2}$, $\varepsilon^{\ph}_{1} + U$, and $\varepsilon^{\ph}_{2} + U$ transport channels open. In purely electronic systems without a vibrational mode, the current plateaus between steps, but in Figs. (\ref{fig: 2a}) and (\ref{fig: 2c}), the current increases gradually with the voltage. This is due to the interaction with the vibrational mode, which, since it has such a low $\omega$ and $\lambda$, can exchange an essentially continuous amount of energy with the electronic degrees of freedom.

These steps are also visible in Figs. (\ref{fig: 2b}) and (\ref{fig: 2d}), which contain the corresponding steady-state vibrational excitation. At low voltages, only the first electronic transport channel is open and $\langle N \rangle^{\text{ss}}$ increases with voltage in a similar manner to the 1L1M model shown in Fig. (\ref{fig: 1d}). As the second transport channel opens, at $e\Phi = 2\varepsilon^{\ph}_{2} = 1.2\text{ eV}$, the vibrational excitation starts decreasing with increasing voltage. Such local cooling arises from a competition between the already open transport channel and the channel that is currently opening, and has been reported for a similar system in Ref.\ \cite{Haertle2011a}. When only the first transport channel is open, increasing the bias voltage increases the excess electronic energy available to be transferred to the vibrational mode, hence the current-induced excitation. As the second transport channel opens, high-energy electrons can also transport almost adiabatically through $\varepsilon^{\ph}_{2}$, exchanging only a small amount of energy with the vibrational mode, hence the decrease in vibrational excitation. This process continues as the four transport channels successively start to open and then saturate. There is a particularly large peak in $\langle N \rangle^{\text{ss}}$ between the opening of $\varepsilon^{\ph}_{2}$ and $\varepsilon^{\ph}_{1} + U$ because the strong Coulomb repulsion induces a large energetic gap between these charging energies. If one were to increase the voltage beyond $\Phi = 5\text{ V}$, then $\langle N \rangle^{\text{ss}}$ would increase monotonically, as all transport channels are fully open. The electron-electron interaction, therefore, has an overall stabilizing effect on the nanojunction in comparison to the one-level, one-mode model. 

Finally, note that in this interacting system, the size of the molecule-lead coupling in comparison to the vibrational frequency is crucial for the accuracy of the semiclassical HEOM-LD approach. In the top row, where $\Gamma < \omega$, the timescale separation assumption is violated, and the HEOM-LD approach clearly does not approximate the steady-state observables as well as it does in the bottom row, where $\Gamma > \omega$. This contrasts the steady-state observables of the one-level, one-mode system, where the HEOM-LD approach performed well even outside the timescale separation assumption. The key difference is that the electron-electron interaction slows electronic relaxation, such that the molecule-lead coupling must be larger to maintain a Markovian approximation of the electronic forces.

\subsection{Influence of a Quantum Electronic-Vibrational Interaction} \label{subsec: One electronic level coupled to two vibrational modes with different timescales}

In a previous publication, two of the authors investigated the electronic friction of a system containing a single electronic level coupled to both a low-frequency nuclear mode, which could be treated in the semiclassical HEOM-LD framework, and a high-frequency nuclear mode, which must be treated quantum mechanically \cite{Rudge2023}. For brevity, this model will be referred to as the one-level, two-mode (1L2M) in the following. The motivation for such a model could be a molecule with some high-frequency internal vibration accompanied by low-frequency center-of-mass motion, or coupling of the molecule to a cavity mode \cite{Chen2019b}. The purpose of this subsection is to explore how the addition of the quantum high-frequency mode affects the dynamics of the classical low-frequency mode.

\begin{center}
\begin{figure}
\centering
\begin{subfigure}[b]{0.0\textwidth}
\phantomcaption\label{fig: 3a}
\end{subfigure}
\begin{subfigure}[b]{0.0\textwidth}
\phantomcaption\label{fig: 3b}
\end{subfigure}
\begin{subfigure}[b]{0.0\textwidth}
\phantomcaption\label{fig: 3c}
\end{subfigure}
\begin{centering}
\includegraphics[width = \columnwidth]{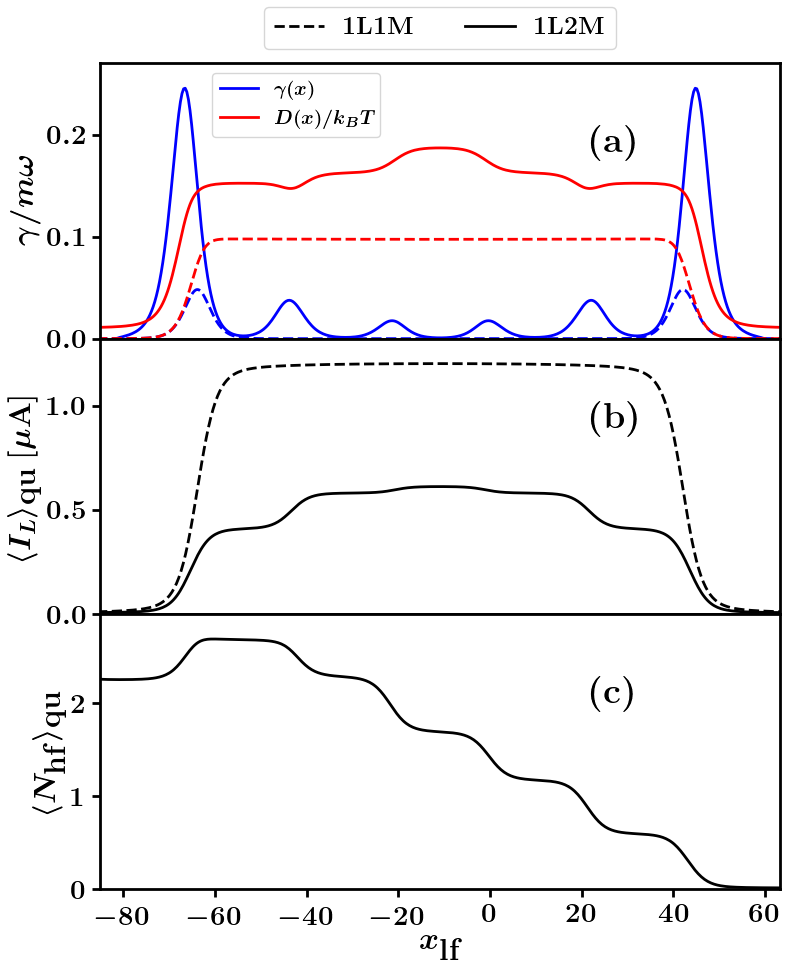}
\caption{Electronic friction and correlation function of the stochastic force (a), instantaneous steady-state electronic current in the adiabatic approximation (b), and vibrational excitation of the high-frequency mode (c), as a function of the low-frequency vibrational coordinate and at finite bias voltage: $\Phi = 1.5\text{ V}$. Solid lines correspond to the 1L2M model and dashed lines correspond to the 1L1M model evaluated at the same parameters but without the high-frequency mode. Parameters are $\Gamma = 20\text{ meV}$, $k_{B}T = 25.8\text{ meV}$, $\omega_{\text{lf}} = 30\text{ meV}$, $\lambda_{\text{lf}} = 10\text{ meV}$, $\omega_{\text{hf}} = 300\text{ meV}$, $\lambda_{\text{hf}} = 450\text{ meV}$. Again, the electronic energy level is chosen such that the energy after the small polaron shift(s) is $\varepsilon_{0} = 150 \text{ meV}$.}
\label{fig: 3}
\end{centering}
\end{figure}
\end{center}

The molecular Hamiltonian is 
\begin{align}
H_{\text{mol}} & = \left(\tilde{\varepsilon}_{0} + \lambda_{\text{lf}}\sqrt{2}\:\hat{x}_{\text{lf}} + \lambda_{\text{hf}}\sqrt{2}\:\hat{x}_{\text{hf}}\right) d^{\dag}_{}d^{\ph} + \nonumber \\
& \:\:\:\:\:\: \frac{\omega_{\text{lf}}}{2}\left(\hat{p}^{2}_{\text{lf}} + \hat{x}^{2}_{\text{lf}}\right) +  \frac{\omega_{\text{hf}}}{2}\left(\hat{p}^{2}_{\text{hf}} + \hat{x}^{2}_{\text{hf}}\right), \label{eq: general system Hamiltonian}
\end{align}
where the ``$\text{lf}$" and ``$\text{hf}$" indices represent low- and high-frequency, respectively. In what follows, only the low-frequency mode will be treated classically, such that $(\hat{p}_{\text{lf}},\hat{x}_{\text{lf}}) \rightarrow (p_{\text{lf}},x_{\text{lf}})$, while the high-frequency mode will be treated quantum mechanically alongside the electrons. In a similar manner, two modes operating on such different timescales require a unique treatment in the fully quantum HEOM, in that the low-frequency mode will be treated as an extra reservoir while the high-frequency mode remains within $\rho(t)$: this is the splitting outlined in Eqs.\eqref{eq: reservoir treatment molecular Hamiltonian}-\eqref{eq: reservoir treatment molecular-lead Hamiltonian}. Note that, to satisfy the requirement of fast relaxation within the quantum part of the system, the electronic-vibrational coupling of the high-frequency mode is set quite large: $\lambda_{\text{hf}} = 1.5\omega_{\text{hf}} = 450\text{ meV}$.

In order to investigate how the addition of the high-frequency mode affects the stability of the low-frequency mode, $\omega_{\text{lf}}$ and $\lambda_{\text{lf}}$ are set to $30\text{ meV}$ and $10 \text{ meV}$, respectively, which corresponds to the green lines in Figs.\ (\ref{fig: 1c}) and (\ref{fig: 1d}). The high-frequency mode is set to $\omega_{\text{hf}} = 300 \text{ meV}$, which, from Figs.\ (\ref{fig: 1e}) and (\ref{fig: 1f}), clearly needs to be treated quantum mechanically. The high-frequency mode is coupled strongly to the electronic level, $\lambda_{\text{hf}} = 450 \text{ meV}$. The energy level is chosen such that $\varepsilon_{0} = \tilde{\varepsilon}_{0} - \frac{\lambda_{\text{lf}}^{2}}{\omega_{\text{lf}}} - \frac{\lambda_{\text{hf}}^{2}}{\omega_{\text{hf}}} = 150 \text{ meV}$. 

In Fig.\ (\ref{fig: 3}), the electronic forces acting on the low-frequency mode and adiabatic quantum transport observables are plotted for the 1L2M model at one example voltage. Shown alongside are also the electronic forces and adiabatic transport observables for the 1L1M model with the same parameters, just without the high-frequency mode. As reported in Ref.\ \cite{Rudge2023},  the electronic friction acting on the low-frequency mode is generally much stronger than in the one-mode system, as there are now extra electron-hole pair creation and annihilation processes via the exchange of energy with the high-frequency mode. These extra processes are also the origin for the extra peaks in the friction for the 1L2M model at $\varepsilon(x_{\text{lf}}) + \nu\omega_{\text{hf}} = e\Phi/2$, where $\varepsilon(x_{\text{lf}}) = \varepsilon - \lambda_{\text{lf}}\sqrt{2}\: x_{\text{lf}}$ and $\nu \in \mathbb{N}$. Since the electronic forces are plotted for a finite bias voltage, the fluctuation-dissipation theorem is not satisfied and the correlation function of the stochastic force displays a broad peak when the level sits within the bias window while the friction does not, which is a symptom of current-induced heating.

\begin{figure}
\begin{subfigure}[b]{0.0\textwidth}
\phantomcaption\label{fig: 4a}
\end{subfigure}
\begin{subfigure}[b]{0.0\textwidth}
\phantomcaption\label{fig: 4b}
\end{subfigure}
\begin{subfigure}[b]{0.0\textwidth}
\phantomcaption\label{fig: 4c}
\end{subfigure}
\begin{centering}
\includegraphics[width = \columnwidth]{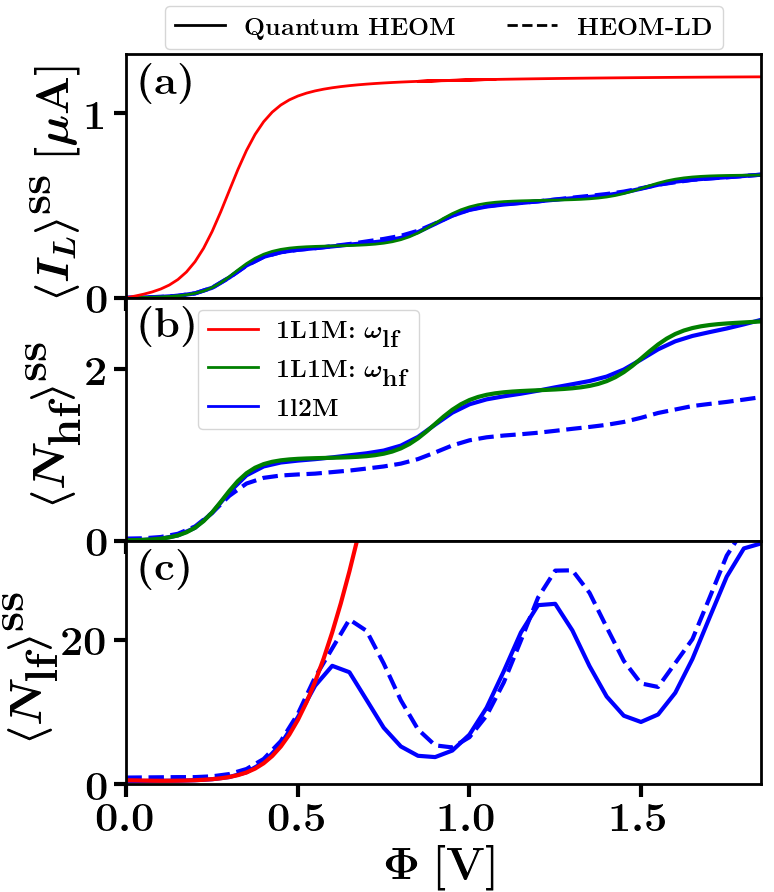}
\caption{Blue: steady state transport observables for the 1L2M model as a function of bias voltage. Electric current is shown in (a), excitation of the high-frequency mode in (b), and excitation of the low-frequency mode in (c). Results calculated from the fully quantum HEOM are given solid lines, while those calculated from the HEOM-LD approach are given dashed lines. Also plotted are the corresponding transport observables for the corresponding single-mode model with just the low-frequency (red) and just the high-frequency (green) mode. All parameters are the same as in Fig.\ (\ref{fig: 3})}
\label{fig: 4}
\end{centering}
\end{figure}

In addition to the extra structure in the electronic friction and correlation function, the instantaneous steady-state electric current also exhibits typical features of Franck-Condon blockade. As shown in Fig.\ (\ref{fig: 3b}), instead of a single plateau when $\varepsilon(x_{\text{lf}})$ lies within the bias window, the current exhibits steps at the same voltages as peaks in the friction and plateaus inbetween. Likewise, the instantaneous steady-state excitation of the high-frequency mode, $\langle N \rangle_{\text{qu}}(x_{\text{lf}})$, displays similar steplike behavior. As the classical coordinate decreases and the level is pushed below the bias window, the level becomes totally occupied and $\langle N \rangle_{\text{qu}}(x_{\text{lf}})$ reaches some saturation value. Evidently, introducing a high-frequency mode to the problem significantly changes the electronic forces and adiabatic transport observables. A natural question, therefore, is how the additional structure shown in Fig.\ (\ref{fig: 3}) manifests itself in the steady state transport observables, which are shown in Fig.\ (\ref{fig: 4}).

First, note that $\langle I_{L} \rangle^{\text{ss}}$ displays steps at the emergence of each phonon-assisted transport channel, $e\Phi/2 = \nu \omega_{\text{hf}}$. This is quantitatively and qualitatively similar behavior to what a single-mode model with $\omega_{\text{hf}}$ and $\lambda_{\text{hf}}$ predicts, which is shown by the solid green line. This is due to the relative strengths of the electronic-vibrational interactions, which differ by an order of magnitude between the high- and low-frequency modes. Unlike the single-mode model, however, $\langle I_{L} \rangle^{\text{ss}}$ does not plateau between steps, instead increasing monotonically due to the essentially continuous exchange of energy with the low-frequency mode. On the other hand, the single-mode model containing $\omega_{\text{lf}}$ and $\lambda_{\text{lf}}$ predicts only one step in the current and saturates to a much higher value, as the suppression of the current due to the Franck-Condon blockade is not present.

A similar difference between the two models can be seen in Fig.\ (\ref{fig: 4b}), in which the excitation of the high-frequency mode is plotted against the voltage. In comparison to the single-mode case, the 1L2M model does not plateau between steps of $\langle N_{\text{hf}}\rangle^{\text{ss}}$. At high bias voltages, the excitation is lower for the 1L2M model, as there is now competition for the energy of transporting electrons with the low-frequency mode. Since $\lambda_{\text{hf}} \gg \lambda_{\text{lf}}$, the excitation is reduced by only a small amount. This is in stark contrast to the excitation of the low-frequency mode, which is shown in Fig.\ (\ref{fig: 4c}). 

At low voltages, $\Phi < 0.6 \text{ V}$, electrons do not have enough incident energy to interact efficiently with the high-frequency mode, and the resulting excitation of the low-frequency mode, $\langle N \rangle_{\text{lf}}^{\text{ss}}$, follows the same monotonic increase that the corresponding 1L1M model with $\omega_{\text{lf}}$ and $\lambda_{\text{lf}}$ displays. After some threshold voltage, $\langle N \rangle_{\text{lf}}^{\text{ss}}$ exhibits negative differential vibrational excitation, with maxima just before the phonon-assisted transport channel at $e\Phi/2 = \nu\omega_{\text{hf}}$ becomes available, and minima inbetween. When these channels are fully open, energy dissipated in the junction is distributed unevenly between the two vibrations due to the large difference in $\lambda_{\text{lf}}$ and $\lambda_{\text{hf}}$, which strongly favors excitation of the high-frequency mode. Adding a strong quantum electronic-vibrational interaction, therefore, has a similar local cooling effect to that already shown in Sec.\ \ref{subsec: Adding an Electron-Electron interaction}.

As a final comment, one should be careful when using the semiclassical approach with additional quantum modes, as the timescale of electronic motion can be strongly affected by the quantum electronic-vibrational coupling. A strong $\lambda_ {\text{hf}}$, for example, leads to the formation of a polaron and an effective molecule-lead coupling of $\Gamma_{\text{eff}} = \Gamma e^{-(\lambda/\omega)^{2}}$ that is orders of magnitude smaller than $\omega_{\text{lf}}$. At room temperature, the HEOM-LD approach is still reasonably accurate in comparison to the full quantum HEOM, and it can always be improved by increasing the molecule-lead coupling. Increasing the molecule-lead coupling in the quantum HEOM reservoir approach, on the other hand, is much more difficult, as one needs more electronic tiers within the hierarchy and calculations for the 1L2M model become prohibitively expensive. 

This speaks to the overall usefulness of the HEOM-LD approach, in that it works best in regimes where it is most difficult to apply the quantum HEOM. Another key restriction to the quantum HEOM approach is that the vibrational degrees of freedom have been modeled as harmonic oscillators with linear electronic-vibrational couplings, which enables the highly efficient reservoir treatment to be used for small $\lambda$. Since realistic molecular models include anharmonic modes and nonlinear electronic-vibrational couplings, for which such an approach \textit{cannot} be used, regimes of high vibrational excitation become very difficult to model. In a discrete variable representation (DVR), for example, large amplitude motion or the motion of heavy nuclei requires many grid points and, correspondingly, a large basis of vibrational states in $\rho(t)$ \cite{Ke2022}. In these scenarios, mixed quantum-classical approaches are significantly more efficient than their quantum counterparts. The HEOM-LD approach, furthermore, performs well in these regimes even when the quantum part of the nanosystem contains strong interactions.

\section{Conclusion} \label{sec: Conclusion}

This work introduced the HEOM-LD approach as a novel mixed quantum-classical method for simulating the nonadiabatic dynamics of molecules under the influence of quantum electronic forces from one or more metal surfaces. The method combines the numerically exact HEOM approach for electronic degrees of freedom with a Langevin equation for the vibrational degrees of freedom within the molecule. Within the new approach, strong interactions within the quantum part of the system can be treated while simultaneously incorporating nonadiabatic effects in the vibrational dynamics, such as relaxation due to electron-hole pair creation. 

To demonstrate the efficacy of HEOM-LD, steady-state transport observables for a molecular nanojunction model were presented, alongside exact calculations from the fully quantum HEOM approach. For a simple vibronic model, where the electronic part of the system is noninteracting, HEOM-LD performs excellently even when the timescale separation is formally violated. Only in regimes where vibrational quantization becomes important, $k_{B}T \ll \omega$, are there differences between the quantum and HEOM-LD methods. Under the influence of an electron-electron or quantum electronic-vibrational interaction, however, the electronic relaxation is slowed and fulfilling the timescale separation assumption is crucial for accurate dynamics. Such strong interactions also introduce a stabilizing effect to the low-frequency mode, which manifests via negative differential vibrational excitation. 

These calculations highlight the strong potential of HEOM-LD as a method for modeling the nonadiabatic dynamics of molecules interacting with metal surfaces, with a particular focus on systems containing strong interactions. Possible applications include investigating the influence of the electron-electron interaction on the desorption or scattering of a molecule from a metal surface, light-induced processes, and vibrational instabilities in molecular nanojunctions arising from nonconservative electronic forces. 

\section{Acknowledgements}\label{sec: Acknowledgements}

This work has been supported by the Deutsche Forschungsgemeinschaft (DFG) within the framework of the Research Unit FOR5099 ``Reducing complexity of nonequilibrium systems'' with Project No. 431945604. S.L.R thanks the Alexander von Humboldt Foundation for the award of a Research Fellowship. Furthermore, support by the state of Baden-W\"{u}rttemberg through bwHPC and the DFG through Grant No. INST 40/575-1 FUGG (JUSTUS 2 cluster) is gratefully acknowledged.

%
%\section{Appendix}
%
%\subsection{Limiting behavior of high-frequency vibrational excitation in 1L2M model} \label{app: Limiting behavior of high-frequency vibrational excitation in 1L2M model}

\bibliography{Main_text_incl._fig.bib} 

\end{document}